\documentstyle[preprint,aps]{revtex}
\begin{document}
\preprint{OITS-548}
\draft
\title{ PROTON DECAY IN SUPERSYMMETRIC FINITE GRAND UNIFICATIONE\footnote{Talk
presented by X.-G. He  at the Eighth Meeting the American Physical Society,
Division of Particles and
Fields (DPF'94), Albuqurque, New Mexico, August 2-6, 1994}}
\author{N.DESHPANDE, XIAO-GANG HE, AND E. KEITH}
\address{Institute of Theoretical Science\\ University of Oregon\\ Eugene,
OR 97403, USA}
 \date{August, 1994}
\maketitle
\begin{abstract}
 We study proton decay in finite supersymmetric
SU(5) grand unified theories. We
find that the finite supersymmetric SU(5) models are ruled out from this
consideration.
\end{abstract}
\newpage

Proton decays are predicted in many grand unified theories (GUTs)\cite{gut}.
Experimentally
no proton decays have been observed\cite{bound}. The stringent experimental
bounds on proton
decays can provide interesting constraints on GUTs\cite{pd,ano,hisano}. It has
been shown that in the
minimal supersymmetric (SUSY) SU(5) model, a large region in parameter space
can be ruled out from this consideration\cite{ano,hisano}.
Here we show that a class of finite SUSY SU(5) model is ruled out by
experimental
bounds on the proton life-time\cite{he}.

There have been many studies of finite
GUTs\cite{discrete,raby}. This is a class of interesting
GUTs. It supports strongly the hope that the ultimate theory does not need
infinite renormalization.  In order to have a finite theory to all orders, the
$\beta$ functions for the gauge coupling and Yukawa couplings have to be zero
to all orders.
The requirement that the $\beta$ function of the gauge coupling be zero greatly
restricts the allowed matter representations in a
theory. $\beta=0 $ for the Yukawa
couplings can put additional constraints on the theory. A
particularly interesting class of theories are the ones based on the
$SU(5)$ gauge group with supersymmetry.
If one requires that $SU(5)$ is broken by the Higgs mechanism to $SU(3)_C\times
SU(2)_L\times U(1)_Y$ with three generations of matter fields,  only one
solution is allowed with
$5$, $\bar 5$, $10$, $\bar {10}$ and  $24$ chiral multiplets with
multiplicities (4,7,3,0,1)\cite{discrete,raby}. This model contains one 24
($\Sigma$) of
Higgs for the SU(5) breaking, $4(5 +\bar 5)$ $(H_\alpha\;,\; \bar H_{\alpha})$
of Higgs some of which will be used for electroweak breaking and the remaining
$3(\bar 5 +10)$ are identified with
the three generation matter fields. With this content, the most general
superpotential that may be written, consistent with renormalizibility, $SU(5)$
invariance and R-parity conservation is of the form
\begin{eqnarray}
W = q Tr\Sigma^3 + M Tr\Sigma^2 + \lambda_{\alpha\beta} \bar H_\alpha \Sigma
H_{\beta} + m_{\alpha\beta}\bar H_\alpha H_\beta + {1\over 2}g_{ij\alpha}
10_i 10_j H_\alpha + \bar g_{ij\alpha} 10_i \bar 5_j \bar H_\alpha\;,
\end{eqnarray}
The indices $\alpha,\; \beta$, and $i,\;j$ run from 1 to 4 and 1 to 3,
respectively.

The requirement that the $\beta$ functions for the Yukawa couplings are zero at
the one-loop level implies,
\begin{eqnarray}
\Sigma: &{189\over 5}q^2 = 10g^2 -
\lambda_{\alpha\beta}\lambda^{\alpha\beta}\;,\nonumber\\
\bar H_\alpha: & \bar g_{ij\alpha} \bar g^{ij\beta} = {6\over 5}
(g^2\delta^\beta_\alpha - \lambda_{\alpha\gamma}\lambda^{\beta\gamma})\;,
\nonumber\\
\bar 5_i: & \bar g_{ki\alpha}\bar g^{kj\alpha} = {6\over 5} g^2\delta^j_i\;,\\
H_\alpha: & g_{ij\alpha} g^{ij\beta} = {8\over 5}(g^2\delta^\beta_\alpha -
\lambda_{\gamma\alpha}\lambda^{\gamma\beta})\;,\nonumber\\
10_i: & 2 g_{ik\alpha} g^{jk\alpha} + 3 g_{ik\alpha} g^{jk\alpha} = {36\over 5}
g^2\delta^j_i\;.\nonumber
\end{eqnarray}
Imposing an additional $Z_7\times Z_3$ symmetry\cite{discrete}, one obtains a
unique solution to eq.(2):
\begin{eqnarray}
g^2_{111} = g^2_{222} = g^2_{333} = {8\over 5}g^2\;,
\;\bar g^2_{111} = \bar g^2_{222} = \bar g^2_{333} = {6\over 5}g^2\;,
\;\lambda_{44}=g^2\;,\;\;\;\;\; q^2 = {5\over 21}g^2\;.
\end{eqnarray}
All other tri-linear couplings are zero.

In the above model only $H_4 (\bar H_4)$ can develop vacuum expectation values
in order that the doublet-triplet mass splitting is possible for the doublets
which
break $SU(3)_C\times SU(2)_L\times U(1)_Y$ to $U(1)_{em}$. It is possible to
find solutions
such that each Higgs doublet can develop a vacuum expectation value and at the
same time it is still possible to maintain the doublet-triplet mass splitting
if
the discrete symmetry is softly broken.
All fermions can have masses\cite{raby}. Carrying out the
renormalization group analysis from GUT scale to the electroweak scale,
the top quark mass is found to be between $175$ to $190$
GeV\cite{discrete}. This is a very interesting prediciton.

There are, however, several problems with this model. Because the Yukawa
couplings are
diagonal, all KM angles are zero. This
problem can be solved by abandoning the
diagonal solution to eq.(2). It is possible to find a solution of eq.(2)
such that
KM matrix can be reproduced. A possible solution is
\begin{eqnarray}
\bar g_{ij\alpha} = \sqrt{6\over5}g(\delta_{i,1}\delta_{\alpha,1}
+\delta_{i,2}\delta_{\alpha,2}+\delta_{i,3}\delta_{\alpha,3})V_{ij}
\end{eqnarray}
with all other couplings the same as in eq.(3). Here $V_{ij}$ is the KM matrix.
This model
has the same predictions for the quark masses.

This model also predicts
the wrong mass relations for the first two generations:
$m_e=m_d$, $m_\mu = m_s$ at the GUT scale  because there are only $5$ and $\bar
5$
Higgs representations to generate masses for quarks and charged leptons.
 If higher dimension operators
are somehow allowed, this problem can be solved. For example,
adding a $(10\times \bar 5)
(\Sigma \bar H_\alpha)$ term can correct the wrong mass relations. This,
however, is not the
major
problem. In the following we will show that even if we relax the conditions to
allow the above additions
to the theory, the model has another problem. It predicts too rapid a proton
decay.

There are several mechanisms by which proton decays
may be induced in SUSY SU(5) theories.
The exchanges of heavy gauge bosons, exchange
of scalar color triplets, and dimension-five operator induced by exchange
s-particles can all lead to proton decays.  The
most significant contributions to the proton decays come from the
dimension-five operator induced by exchanging color triplet higgsinos $H_C$ and
$\bar H_C$ of
$H_\alpha$ and $\bar H_\alpha$\cite{pd,ano,hisano} and wino in the loop. In the
minimal SUSY SU(5)
model, this mechanism is the dominant one and considerably restricts
the allowed region in parameter space\cite{hisano}. In the
finite SU(5) model, experimental bounds on proton decays all but make these
models unacceptable\cite{he}. The four-fermion baryon number violating
effective Lagrangian
at 1 GeV can be written down explicitly as\cite{hisano}
\begin{eqnarray}
L &=& {\alpha_2\over 2\pi M_{H_C^\alpha}} g_{ii\alpha} \bar g_{kk\alpha}
V_{jk}^*A_SA_L\nonumber\\
&&\times [ (u_id'_i)(d'_j\nu_k) ( f(u_j,e_k) + f(u_i, d'_i))+
(d'_iu_i)(u_je_k)(f(u_i,d_i)+f(d_j', \nu_k))\\
&&+ (d'_i\nu_k)(d'_iu_j)(f(u_i, e_k)+f(u_i, d'_j))+(u_id'_j)(u_ie_k) (f(d'_i,
u_j)+f(d'_i,\nu_k))]\nonumber\;,
\end{eqnarray}
where $d'_i = V_{il}d_l$; $f(a,b)= m_{\tilde w}[m^2_{\tilde a} \ln (m^2_{\tilde
a}/m^2_{\tilde w})/(m^2_{\tilde a}-m^2_{\tilde w}) - (m_{\tilde a} \rightarrow
m_{\tilde b})]/( m^2_{\tilde a}-m^2_{\tilde b})$ is from the loop integral, and
$m_{\tilde a, \tilde b}$ are the s-fermion masses,
$A_S \approx 0.59$, $A_L \approx 0.22$\cite{hisano} are the QCD correction
factors for the
running from $M_{GUT}$ to SUSY breaking scale and from SUSY breaking scale to 1
GeV, respectively, and the Yukawa couplings are evaluated at 1 GeV.

Because all $g_{ii\alpha}$ and $\bar g_{jj\alpha}$ are equal in the model we
are considering, the dominant contributions to the proton decays
will be the ones involving only particles in the first generation. The dominant
baryon number violating decay modes are:
$p\rightarrow \pi^+ \bar \nu_e$, $p\rightarrow \pi^0 (\eta) e^+$,
$n\rightarrow \pi^0 (\eta) \bar \nu_e$, $p\rightarrow \pi^- e^+$.

Finally to
obtain the life times of the proton and neutron, we employ the chiral
Lagrangain
approach to parametrize the hadronic matrix elements.
We have
\begin{eqnarray}
\Gamma&(&p\rightarrow \pi^+ \bar \nu_e) = 2\Gamma(n\rightarrow \pi^0 \bar
\nu_e)
= \beta^2 {m_N\over 32\pi f^2_{\pi}}|C(duu\nu_e)(1+D+F)|^2\nonumber\;,\\
\Gamma&(&n\rightarrow \eta \bar \nu_e)
=\beta^2 {(m_N^2-m_\eta^2)^2\over 64\pi f^2_{\pi}m_N^3}3|C(duu\nu_e)(1-{1\over
3}(D-3F))|^2\;,\\
\Gamma&(&n\rightarrow \pi^- e^+) = 2\Gamma(p\rightarrow \pi^0 e^+)
= \beta^2 {m_N\over 32\pi f^2_{\pi}}|C(duue)(1+D+F)|^2\nonumber\;,\\
\Gamma&(&p\rightarrow \eta e^+)
=\beta^2 {(m_N^2-m_\eta^2)^2\over 64\pi f^2_{\pi}m_N^3}3|C(duue)(1-{1\over
3}(D-3F))|^2\;,\nonumber
\end{eqnarray}
where $D = 0.81$ and $F=0.44$, which arise from the strong interacting
baryon-meson
chiral Lagrangian, $f_\pi = 132$ MeV is the pion decay constant, and $m_N$ and
$m_\eta$ are the neucleon and $\eta$ meson masses, respectively. The parameter
$\beta$ is estimated to be in the range\cite{beta} $0.03$ GeV$^3$ to $0.0056$
GeV$^3$. The parameters
$C(duu\nu)$ and $C(duue)$ are the coefficients of the operators
$(du)(u\nu)$ and $(du)(ue)$ which can be read off from eq.(5). We have
\begin{eqnarray}
C(duu\nu_e) &=& {4\alpha^2_{em}\over \sin^4\theta_W}{\bar m_b \bar m_t\over
m_W^2 \sin 2\beta_H}{A_SA_L\over
M_{H_C^1}}V_{ud}^2V^*_{ud} (f(u,e)+f(u,d))\;,\nonumber\\
C(duue) &=&{4\alpha^2_{em}\over \sin^4\theta_W}{\bar m_b \bar m_t\over m_W^2
\sin 2\beta_H}{A_SA_L\over
M_{H_C^1}}V_{ud}V^*_{ud} (f(u,e)+f(u,d))\;.
\end{eqnarray}

In the above we have used $g_{111}\bar g_{111} = g_2^2\bar m_b\bar
m_t/m_W^2\sin2\beta_H$ as a good approximation. Here the quark masses are at 1
GeV. $\tan \beta_H$ is the ratio of the vacuum expectation value of $H_1$ to
that of $\bar H_1$. It is predicted to be about 50. The top quark mass at 1 GeV
$\bar m_t$ is about 470 GeV\cite{discrete}. Using these values,
we obtain the partial life-times for some of the baryon number violating decays
as
\begin{eqnarray}
\tau&(&p\rightarrow \pi^0 e^+)\approx \tau(n\rightarrow \pi^0\bar \nu_e)
\approx 6 \times 10^{17}\times P \;\;years\;,\nonumber\\
\tau&(&p\rightarrow \pi^+\bar \nu_e)\approx \tau(n\rightarrow \pi^- e^+)
\approx 3 \times 10^{17}\times P\;\; years\;,\\
\tau&(&p\rightarrow \eta e^+)\approx \tau(n\rightarrow \eta \bar \nu_e)
\approx 2 \times 10^{18}\times P\;\; years\;,\nonumber
\end{eqnarray}
where
\begin{eqnarray}
P = \left ({0.003\;GeV^3\over \beta}\right )^2\left (
{M_{H_C}\over 10^{17}\; GeV}{TeV^{-1}\over f(u,d)+f(u,e)}\right )^2\;.
\end{eqnarray}
Using $\beta = 0.003 GeV^3$, we find that even
if we allow $m_{H_C}$ to be the same order as the Planck mass, these partial
life-times are in contradiction with experiments if the factor
$I=TeV^{-1}/(f(u,d)+
f(u,e))$ is of order one. There are two possible ways this problem can be
avoided.
One requires the wino mass to be larger than $10^8$ TeV. Another forces
 the s-fermion masses to be
much larger than the wino mass. The
s-fermion masses have to be larger than $2\times 10^3$ TeV for $m_{\tilde w} >
100$ GeV with $m_{H_C} = 10^{17}$ GeV.
All these solutions require that SUSY be broken at a scale much much larger
than
a few TeV. However such solutions spoil the nice feature of solving the
hierarchy problem that is the rationale for using SUSY theories in the first
place. From these consideration, the model discussed above is
either ruled out, or quite unattractive needing a large SUSY breaking scale.
We expect this
problem to arise in most finite thoeries of grand unification that allow proton
decay.

This work is supported in part by Department of Energy No. DE-FG06-85ER40224.

\end{document}